\documentclass[a4paper]{article}

\usepackage{INTERSPEECH2020}
\usepackage{breqn}
\usepackage{mathtools}
\usepackage{amsmath}
\usepackage[mathscr]{euscript}
\usepackage{xcolor}
\usepackage{url}

\title{Multi-speaker Emotion Conversion via Latent Variable Regularization \\
and a Chained Encoder-Decoder-Predictor Network}
\name{Ravi Shankar$^1$, Hsi-Wei Hsieh$^2$, Nicolas Charon$^2$, Archana Venkataraman$^1$}
\address{
  $^1$Department of Electrical and Computer Engineering, Johns Hopkins University \\
  $^2$Department of Applied Mathematics and Statistics, Johns Hopkins University}
\email{rshanka3@jhu.edu, \{hsieh,charon\}@cis.jhu.edu, archana.venkataraman@jhu.edu}

\begin{document}

\maketitle
\begin{abstract}
We propose a novel method for emotion conversion in speech based on a chained encoder-decoder-predictor neural network architecture. The encoder constructs a latent embedding of the fundamental frequency (F0) contour and the spectrum, which we regularize using the Large Diffeomorphic Metric Mapping (LDDMM) registration framework. The decoder uses this embedding to predict the modified F0 contour in a target emotional class. Finally, the predictor uses the original spectrum and the modified F0 contour to generate a corresponding target spectrum. Our joint objective function simultaneously optimizes the parameters of three model blocks. We show that our method outperforms the existing state-of-the-art approaches on both, the saliency of emotion conversion and the quality of resynthesized speech. In addition, the LDDMM regularization allows our model to convert phrases that were not present in training, thus providing evidence for out-of-sample generalization. 
\end{abstract}

\section{Introduction}
Automated speech synthesis has radically transformed our interaction with machines. It is used in assistive technologies, such as screen readers for the visually impaired, and hands-free devices, such as Amazon's Echo. Emotional speech synthesis is the next milestone in this domain~\cite{vocal_comm_emotion, psych}. For example, emotional machines can be deployed in call centers, where customer frustration is a regular occurrence, and it can provide a better foundation for virtual companions for the elderly or impaired. 
\par
The quality of machine-generated speech has improved phenomenally in the last decade, largely due to the representational power of deep neural networks~\cite{wavenet, tacotron, lpcnet}, which are trained on hundreds of hours of transcribed human speech. However, controlling the expressiveness of synthetic speech remains an open challenge. Recent works in emotional speech synthesis include~\cite{mellotron}, which generates singing voice conditioned on the input rhythm, pitch and linguistic features. A disentangled model for style and content is proposed by~\cite{latent_uncovering, latent_uncovering_2} to infer the latent representations responsible for expressiveness. While these models represent seminal contributions to emotional speech synthesis, the latent representations are learned in an unsupervised manner, which makes it difficult for the user to control the output emotion. Another problem is the poor rate of speech generation due to the auto-regressive nature of these models~\cite{survey_deep_speech}. These challenges motivate the study of emotion conversion as an alternative to end-to-end synthesis approaches. Notably, emotion conversion methods provide controllability over the generated affect, they require much less data to train, and the processing speed is high enough for real-time applications.
\par
Several interesting approaches for emotion conversion have been proposed in the recent past. For example, the work of~\cite{gmm_emo_conv} uses a Gaussian Mixture Model with global variance constraint (GMM-GV) to modify the fundamental frequency (F0) contour and the spectrum. A bidirectional long-short term memory (Bi-LSTM) based architecture has been proposed by~\cite{lstm_emo_conv} to estimate the F0 contour and the spectral features of the target emotion utterance. Another approach by~\cite{hnet_max_likelihood} converts the pitch contour and energy contour of the source utterance using a highway neural network which maximizes the error log likelihood in an expectation-maximization scheme. The same authors further proposed a curve registration based method~\cite{diffeomorphic_hnet} to modify only the F0 contour. Finally, a cycle-consistent generative adversarial network (cycle-GAN) proposed by~\cite{cyclegan_emo_conv} learns to sample the pitch contour and the spectrum from the  target emotional class in an unsupervised manner. While these methods have been successful in single-speaker settings, many of them fail on multispeaker dataset due to the larger overlap of F0 and spectral features between emotional classes. In this paper we propose a novel approach to model the relationship between the F0 contour and the spectral features, deriving it from the basic knowledge of these two representations. Furthermore, unlike other existing methods, our chained estimation also minimizes the mismatch between F0 and the corresponding spectral harmonics. Our second contribution in this paper is to implicitly model the target pitch contour as a smooth and invertible warping of source F0 contour. This is done by learning a latent embedding based on the Large Diffeomorphic Metric Mapping (LDDMM)~\cite{diffeomorphism, landmark_diffeomorphism} framework. In essence the embedding serves as an intermediary between the source and target emotions. We demonstrate that imposing this constraint improves the prediction of the pitch contour significantly. 
\par
Our architecture consists of three separate convolutional neural networks for predicting the embedding, the pitch contour, and the spectrum, respectively. These networks are trained in an end-to-end fashion from a unified objective function. We compare our model against three state-of-the-art baseline methods using the multispeaker VESUS dataset~\cite{vesus}. We further demonstrate that our model does well on sentences, which are not part of the training set, establishing its generalization capability. Finally, in addition to emotion conversion, we show that the proposed model generates better quality of speech than the baselines from both supervised and unsupervised domain. 

\section{Method}
Our novel method uses a chained encoder-decoder-predictor network architecture to modify both the spectrum and the F0 contour of an utterance. The three components of the architecture are jointly optimized through a unified loss function. 
\begin{figure}[!t]
  \centering
  \includegraphics[width=0.98\linewidth, height=3.7cm]{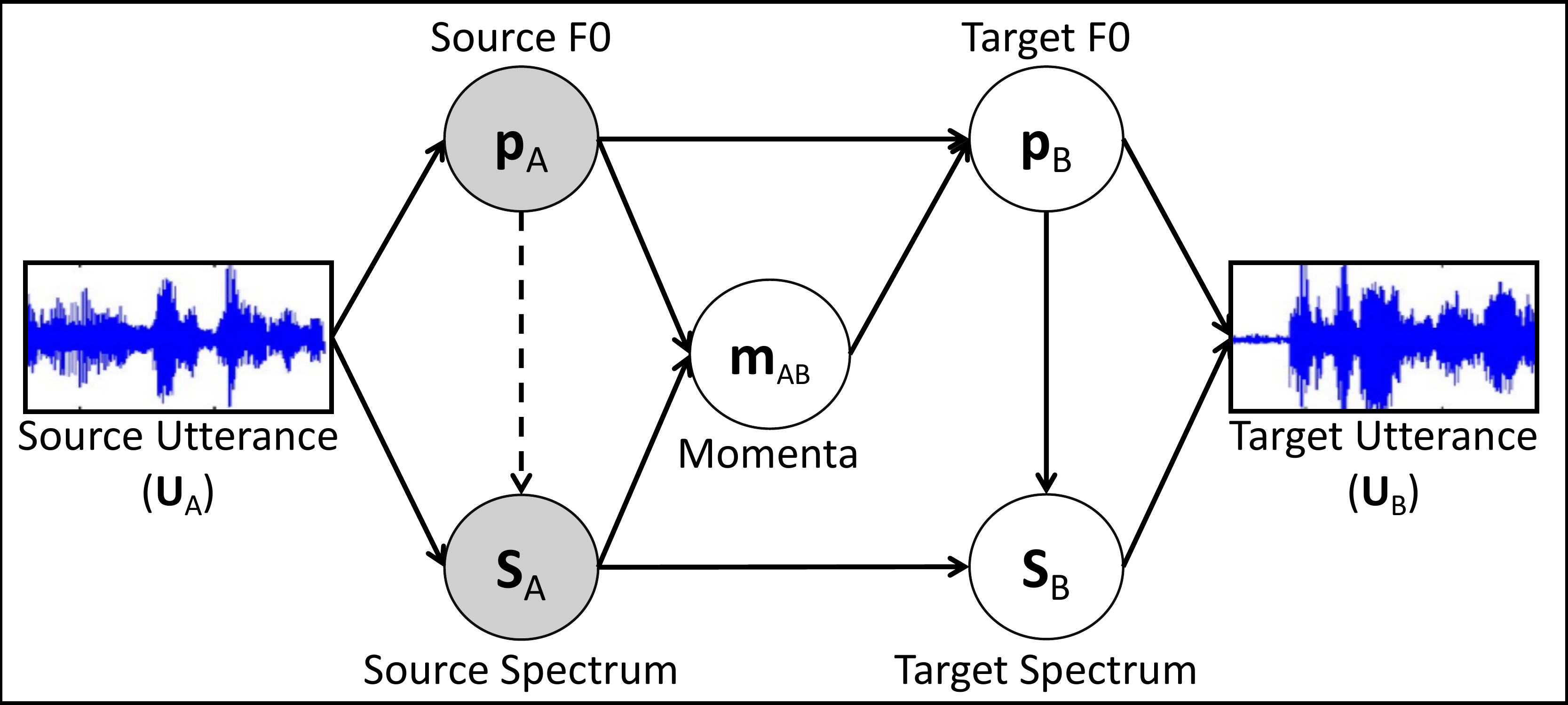}
  \caption{Graphical model of our emotion conversion strategy. $\mathbf{m}_{AB}$ is the intermediary between emotion classes.}
  \label{fig:graphical_model}
\vspace{-5mm}
\end{figure}

Fig.~\ref{fig:graphical_model} describes the relationship between the random variables in our model. We use WORLD vocoder~\cite{straight, world_vocoder} for the analysis and synthesis of speech. Given a source-target pair of emotional utterances denoted by \textbf{U$_{A}$} and \textbf{U$_{B}$}, respectively, the source utterance is decomposed into its components: the spectrum (\textbf{S$_{A}$}) and the F0 contour (\textbf{p$_{A}$}). These components allow us to estimate an intermediate parameter, known as the momenta (\textbf{m$_{AB}$}). From here, the target F0 contour (\textbf{p$_{B}$}) is modeled as a function of the source F0 contour (\textbf{p$_{A}$}) and the momenta (\textbf{m$_{AB}$}). Next, we estimate the target spectrum (\textbf{S$_{B}$}) given the target F0 contour (\textbf{p$_{B}$}) and the source spectrum (\textbf{S$_{A}$}). Finally, the estimated variables are used to synthesize the target emotion utterance. The joint distribution shown in Fig.~\ref{fig:graphical_model} factorizes as:
\vspace{-3.5mm}
\begin{multline}
\text{\scriptsize $ P(\mathbf{p}_{A},\mathbf{S}_{A},\mathbf{m}_{AB},\mathbf{p}_{B},\mathbf{S}_{B}) = P(\mathbf{p}_{A}) \times P(\mathbf{S}_{A}|\mathbf{p}_{A}) $} \\
\text{\scriptsize $ \times P(\mathbf{m}_{AB}|\mathbf{p}_{A},\mathbf{S}_{A}) \times P(\mathbf{p}_{B}|\mathbf{p}_{A},\mathbf{m}_{AB}) \times P(\mathbf{S}_{B}|\mathbf{S}_{A},\mathbf{p}_{B}) $} 
\label{eqn:joint_dist}
\end{multline}
\vspace{-5mm}

\begin{figure*}[!t]
  \centering
  \includegraphics[width=0.95\textwidth, height=5cm]{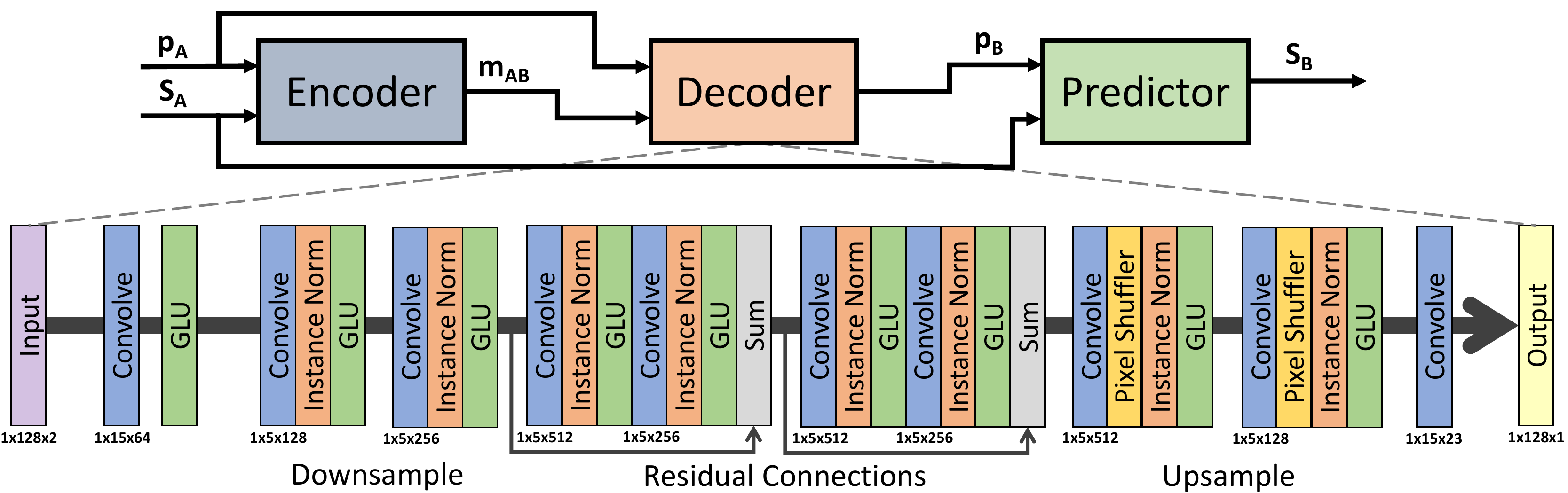}
  \caption{Block model representation of the encoder-decoder-predictor. Encoder and decoder use the same architecture whereas predictor has an extra residual block. GLU in the model stands for the gated linear unit. We use instance normalization due to small mini-batch size and pixel shuffling for up-sampling. The size and number of kernels are indicated below each convolution block.}
  \label{fig:nn_architecture}
  \vspace{-4mm}
\end{figure*}

\vspace{-2.5mm}
\subsection{Regularization via latent representation}
We use an explicit prior on the latent variable to improve the prediction of F0 and spectrum. Specifically, we model the target F0 contour as a smooth and invertible deformation of the source F0 contour. The idea of smooth deformations has been used extensively for images \cite{image_registration}, but here we use it for 2-D curves. Mathematically, let \textbf{$\mathbf{p}_{A}^{\mathbf{t}}$} and \textbf{$\mathbf{p}_{B}^{\mathbf{t}}$} denote a pair of source and target F0 contours, respectively. The variable $\mathbf{t}$ corresponds to the location of the analysis window as it moves across a given speech utterance. The objective of this deformation process is to estimate a series of small vertical displacements $\mathbf{v_{t}(x;s)}$ \cite{diffeomorphism} over frequency and time. The variable $\mathbf{s} \; \epsilon \; [0,1]$ controls the evolution of these small displacements in the discrete setting. The registration problem can thus be formulated as:
\vspace{-2mm}
\begin{equation}
 \text{\footnotesize $ \min_{\mathbf{v} \epsilon V} \frac{1}{2} \int_{0}^{1} \Arrowvert \mathbf{v_{t}(\cdot;s)} \Arrowvert_{V}^{2} ds + \lambda 
\sum_{t=1}^{T} \Arrowvert \phi_{\mathbf{t}}^{\mathbf{v}}(\mathbf{p}_{A}^{\mathbf{t}};1) - 
\mathbf{p}_{B}^{\mathbf{t}} \Arrowvert_{2}^{2} $}
\label{eqn:lddmm_objective}
\vspace{-2mm}
\end{equation}
Here, $\Arrowvert \cdot \Arrowvert_{V}$ denotes the Hilbert norm which is implicitly defined in our case by a Gaussian kernel. The variable $\phi_{\mathbf{t}}^{\mathbf{v}}$ denotes the net displacement field i.e, $\phi_{\mathbf{t}}^{\mathbf{v}} = \int_{0}^{1}\mathbf{v}_{\mathbf{t}}(\cdot;s)ds$. 
\par
Further, it has been theoretically shown in~\cite{shapes_diffeomorphism, momenta_control_diffeomorphism} that the objective in Eq.~(\ref{eqn:lddmm_objective}) can be reformulated in terms of variables $\mathbf{m}_{\mathbf{t}}^0$, known as the initial momenta, according to: 
\vspace{-2.5mm}
\begin{equation}
\text{\footnotesize $\Gamma(\mathbf{m}^{0}) = \frac{1}{2} 
\sum_{i,j=1}^{T}\gamma_{ij}\mathbf{m}_{\mathbf{i}}^{0} \mathbf{m}_{\mathbf{j}}^{0} + \lambda \sum_{t=1}^{T} \Arrowvert \phi_{\mathbf{t}}^{\mathbf{v}}(\mathbf{p}_{A}^{\mathbf{t}};1) - \mathbf{p}_{B}^{\mathbf{t}} \Arrowvert_{2}^{2} $}
\label{eqn:momenta_objective}
\vspace{-1.5mm}
\end{equation}
The variable $\gamma_{ij}$ is an exponential smoothing kernel evaluated on pairs of time points of the source contour $\mathbf{p}_{A}^{\mathbf{t}}$. 
\par
During training, we solve Eq.~(\ref{eqn:momenta_objective}) for every pair of source and target F0 contours to generate the ground truth momenta. This variable summarizes the transformation between emotion pairs. Since the momenta and source F0 contour uniquely specify the transformation, we use it as an intermediary between any given pair of utterances. In comparison, \cite{diffeomorphic_hnet} predicts a momentum for every frame of the pitch contour and then warps it over several iterations specified by variable $\mathbf{s}$. It is a sub-optimal strategy, as there is no temporal coherence constraint in predicting the momenta. Note that we do not have access to the ground truth momenta during testing and run the network in an open loop fashion without intermediate regularization.

\vspace{-1.5mm}
\subsection{Encoder-Decoder-Predictor Network}
Current methods in emotion conversion modify the F0 and spectrum without imposing any explicit relationship between the features. As a result, there are significant residual harmonics present in the spectrum, which results in the poor quality of resynthesised speech. Our approach overcomes this limitation via the conditional relationships modeled in Fig.~\ref{fig:graphical_model}. Here, the conditional spectrum estimate is given by: 
\vspace{-1mm}
\begin{equation}
\text{\footnotesize $\mathbf{\hat{S}}_{B} = \arg\max_{\mathbf{S}_{B}} P(\mathbf{S}_{B} | \mathbf{S}_{A}, \mathbf{p}_{A}) $}
\label{eqn:spect_nn}
\end{equation}
\vspace{-1mm}
Using rules of probability, we can rewrite Eq.~(\ref{eqn:spect_nn}) as:
\begin{align*}
& \text{\scriptsize $ \mathbf{\hat{S}_{B}} = \arg\max_{\mathbf{S}_{B}} \int_{\mathbf{p}_{B}} P(\mathbf{S}_{B}, \mathbf{p}_{B} | \mathbf{S}_{A}, \mathbf{p}_{A}) \; d\mathbf{p}_{B} $} \\
& \text{\scriptsize $ \phantom{\mathbf{\hat{S}_{B}}} = \arg\max_{\mathbf{S}_{B}} \int_{\mathbf{p}_{B}} P(\mathbf{S}_{B} | \mathbf{S}_{A},\mathbf{p}_{B}) P(\mathbf{p}_{B} | \mathbf{S}_{A}, \mathbf{p}_{A}) \; d\mathbf{p}_{B}$} \\
& \text{\scriptsize $ \phantom{\mathbf{\hat{S}_{B}}} = \arg\max_{\mathbf{S}_{B}} \int_{\mathbf{p}_{B}} P(\mathbf{S}_{B} | \mathbf{S}_{A},\mathbf{p}_{B}) \times \int_{\mathbf{m}_{AB}} \! \! \! \! \! P(\mathbf{p}_{B} | \mathbf{m}_{AB}, \mathbf{p}_{A}) $} \\
& \text{\scriptsize $ \phantom{\mathbf{\hat{S}_{B}} = \;\;\;\;\;\;\;\;\;\;\;\;\;\;} \times P(\mathbf{m}_{AB} | \mathbf{S}_{A}, \mathbf{p}_{A}) \; d\mathbf{m}_{AB} \; d\mathbf{p}_{B} $} \\
& \text{\scriptsize $\phantom{\mathbf{\hat{S}_{B}}} = \arg\max_{\mathbf{S}_{B}} \int_{\mathbf{m}_{AB}} \! \! \! \! \! P(\mathbf{m}_{AB} | \mathbf{S}_{A},\mathbf{p}_{A}) \times \int_{\mathbf{p}_{B}} \! \! P(\mathbf{p}_{B} | \mathbf{m}_{AB}, \mathbf{p}_{A}) $} \\
& \text{\scriptsize $ \phantom{\mathbf{\hat{S}_{B}} = \;\;\;\;\;\;\;\;\;\;\;\;\;\;} \times P(\mathbf{S}_{B} | \mathbf{S}_{A}, \mathbf{p}_{B}) \; d\mathbf{p}_{B} \;  d\mathbf{m}_{AB}, $}
\end{align*}
\vspace{-3.5mm}
\par 
\noindent where we have used Eq.~(\ref{eqn:joint_dist}) to derive the above expression. The first term term we encounter is $P(\mathbf{m}_{AB} | \mathbf{S}_{A}, \mathbf{p}_{A})$ which is the probability density of the intermediate latent representation i.e., momenta. It is conditioned on both, the source F0 contour and the spectrum. The second term, $P(\mathbf{p}_{B} | \mathbf{m}_{AB}, \mathbf{p}_{A})$ is the density over the target F0 contour given the momenta and the source F0 contour. Finally, $P(\mathbf{S}_{B} | \mathbf{S}_{A}, \mathbf{p}_{B})$ is the target spectrum conditioned on the target pitch contour and the source spectrum. Note that the expression requires multiple integrations, and is hence, intractable. However, we can make point estimates for each density function using a deep convolutional neural network~\cite{dcgan} (CNN) thereby, allowing us to write:
\vspace{-0.5mm}
\begin{align}
& \text{\footnotesize $ \mathbf{\hat{m}}_{AB} = \arg\max_{\mathbf{m}_{AB}} P(\mathbf{m}_{AB} | \mathbf{S}_{A}, \mathbf{p}_{A} ; \theta_{e}) $} \nonumber \\
& \text{\footnotesize $ \mathbf{\hat{p}}_{B} \; \; \; = \arg\max_{\mathbf{p}_{B}} P(\mathbf{p}_{B} | \mathbf{\hat{m}}_{AB}, \mathbf{p}_{A} ; \theta_{d}) $} \nonumber \\
& \text{\footnotesize $ \mathbf{\hat{S}}_{B} \; \; \; = \arg\max_{\mathbf{S}_{B}} P(\mathbf{S}_{B} | \mathbf{S}_{A}, \mathbf{\hat{p}}_{B} ; \theta_{p}) $} \\
\label{eqn:point_estimates}
\nonumber
\end{align}
\vspace{-8.5mm}
\par
The CNN approximating $P(\mathbf{m}_{AB} | \mathbf{S}_{A}, \mathbf{p}_{A} ; \theta_{e})$ is called an encoder because it distills information about the input data. The CNN modeling $P(\mathbf{p}_{B} | \mathbf{m}_{AB}, \mathbf{p}_{A} ; \theta_{d})$ is called the decoder because it estimates the output pitch from the latent embedding and source pitch contour. The encoder-decoder portion is a basic sequence-to-sequence model for pitch contours. Finally, the CNN modeling $P(\mathbf{S}_{B} | \mathbf{S}_{A}, \mathbf{p}_{B} ; \theta_{p})$ is called a predictor as it generates the spectrum for the converted speech. 
\par
The architecture of these CNNs is shown in Fig.~\ref{fig:nn_architecture}. We adapt the architecture from~\cite{cycle_gan_vc} by reducing the number of residual layers in each block. The entire sequence of three neural networks is trained together from a unified objective. The loss function for optimizing the parameters is given by:
\begin{align}
& \text{\footnotesize $ \mathscr{L} = -\log \Big( P \Big(\mathbf{m}_{AB}, \mathbf{p}_{B}, \mathbf{S}_{B} | \mathbf{S}_{A}, \mathbf{p}_{A} ; \theta_{e}, \theta_{d}, \theta_{p} \Big) \Big) $} \nonumber \\
& \text{\footnotesize $ = \lambda_{e} \Arrowvert \mathbf{\hat{m}}_{AB} - \mathbf{\bar{m}}_{AB} \Arrowvert_{1} + \lambda_{d} \Arrowvert \mathbf{\hat{p}}_{B} - \mathbf{\bar{p}}_{B} \Arrowvert_{1} + \lambda_{p} \Arrowvert \mathbf{\hat{S}}_{B} - \mathbf{\bar{S}}_{B} \Arrowvert_{1} $} \\
\label{eqn:loss_function}
\nonumber
\end{align}
\par
During training, we minimize the negative log likelihood of momenta and the target features with respect to $\theta$. We model the conditional distribution of each variable by Laplace density function. The corresponding ground truths ($\mathbf{\bar{m}}_{AB}, \mathbf{\bar{p}}_{B}, \mathbf{\bar{S}}_{B}$) are used as the mean while the variances are assumed to be constant. This in turn is equivalent to minimizing the mean absolute error of each target variable with an appropriate scaling, defined by $\lambda_{e}$, $\lambda_{d}$ and $\lambda_{p}$, which are the hyperparameters in our model.
\par
One benefit of coupling the neural networks is that the encoder and the decoder become aware of the downstream task of spectrum prediction. We train the neural network~\cite{tensorflow} using Adam optimizer~\cite{adam_optimizer} with a learning rate of 1e-5 and a mini-batch of size one. 23 dimensional MFCC features are used as spectrum representation extracted by an analysis window of length 5ms. During training, the context size is fixed at 640ms which results in dimensionality of $128\times1$ for F0 contour and $128\times23$ for spectrum. The dimensions of momenta are same as the F0 contour. The hyperparameters, $\lambda_{e}$, $\lambda_{d}$ and $\lambda_{p}$ are set to 0.01, 1e-4 and 1e-4, respectively. We do not normalize the input and output features during training to preserve their scale. Code can be downloaded from: \url{https://engineering.jhu.edu/nsa/links/}.

\section{Experiments and Results}
We carry out an ablation study for the momenta $\mathbf{m}_{AB}$ and a qualitative evaluation of emotional salience and quality.

\vspace{-1mm}
\subsection{Emotional Speech Dataset}
We evaluate our algorithm on the VESUS dataset~\cite{vesus} collected at Johns Hopkins University. VESUS contains 250 parallel utterances spoken by 10 actors (gender balanced) in neutral, sad, angry and happy emotional classes. Each spoken utterance has a crowd-sourced emotional saliency rating provided by 10 workers on Amazon Mechanical Turk (AMT). These ratings represent the ratio of workers who correctly identify the intended emotion in a recorded utterance. For robustness, we restrict our experiments to utterances that were correctly and consistently rated as emotional by at least 5 of the 10 AMT workers. As a result, the total number of utterances used are as follows:
\begin{itemize}
    \item \textbf{Neutral to Angry conversion}: 1534 utterances for training, 72 for validation and, 61 for testing.
    \item \textbf{Neutral to Happy conversion}: 790 utterances for training, 43 for validation and, 43 for testing.
    \item \textbf{Neutral to Sad conversion}: 1449 utterances for training, 75 for validation and, 63 for testing.
\end{itemize}
Our subjective evaluation includes both an emotion perception test and, a quality assessment test. These experiments are carried out on Amazon Mechanical Turk (AMT); each pair of speech utterances is rated by 5 workers. The perception test asks the raters to identify the emotion in the converted speech sample, and the quality assessment test asks them to rate the quality of speech sample on a scale of 1 to 5. We include both the neutral and converted utterances to account for the speaker bias. Further, the samples were randomized to mitigate the effects of non-diligent raters and to identify bots.

\vspace{-1.5mm}
\subsection{Baselines}
We compare our encoder-decoder-predictor model to three state-of-the-art baseline methods. The first approach learns a Gaussian mixture model using concatenated source and target features~\cite{gmm_emo_conv}. During inference, a maximum likelihood estimate of target features is made given the source features. A global variance constraint ensures that the estimate is not over-smooth, which is a common problem in joint modeling techniques. 
\par
The second baseline is a Bi-LSTM supervised learning approach~\cite{lstm_emo_conv}. Since Bi-LSTMs generally require considerable data to train, we adopt the strategy in~\cite{lstm_emo_conv} of training the model on a voice conversion task~\cite{cmu_arctic} and then fine-tuning it for emotion conversion. This method encodes the prosody features via a Wavelet transform to represent both short-term and long-term trajectory information of F0 and energy contours. 
\par
The third baseline is a recently proposed unsupervised method for emotion conversion~\cite{cyclegan_emo_conv}. This algorithm uses cycle-GANs to inject emotion into neutral utterances. A set of cycle-GAN transforms the spectrum while the other set transforms the prosody features. Once again, prosodic features are parameterized using Wavelet basis similar to the Bi-LSTM. 

\begin{figure}[!t]
  \centering
  \includegraphics[width=0.9\linewidth, height=3.5cm]{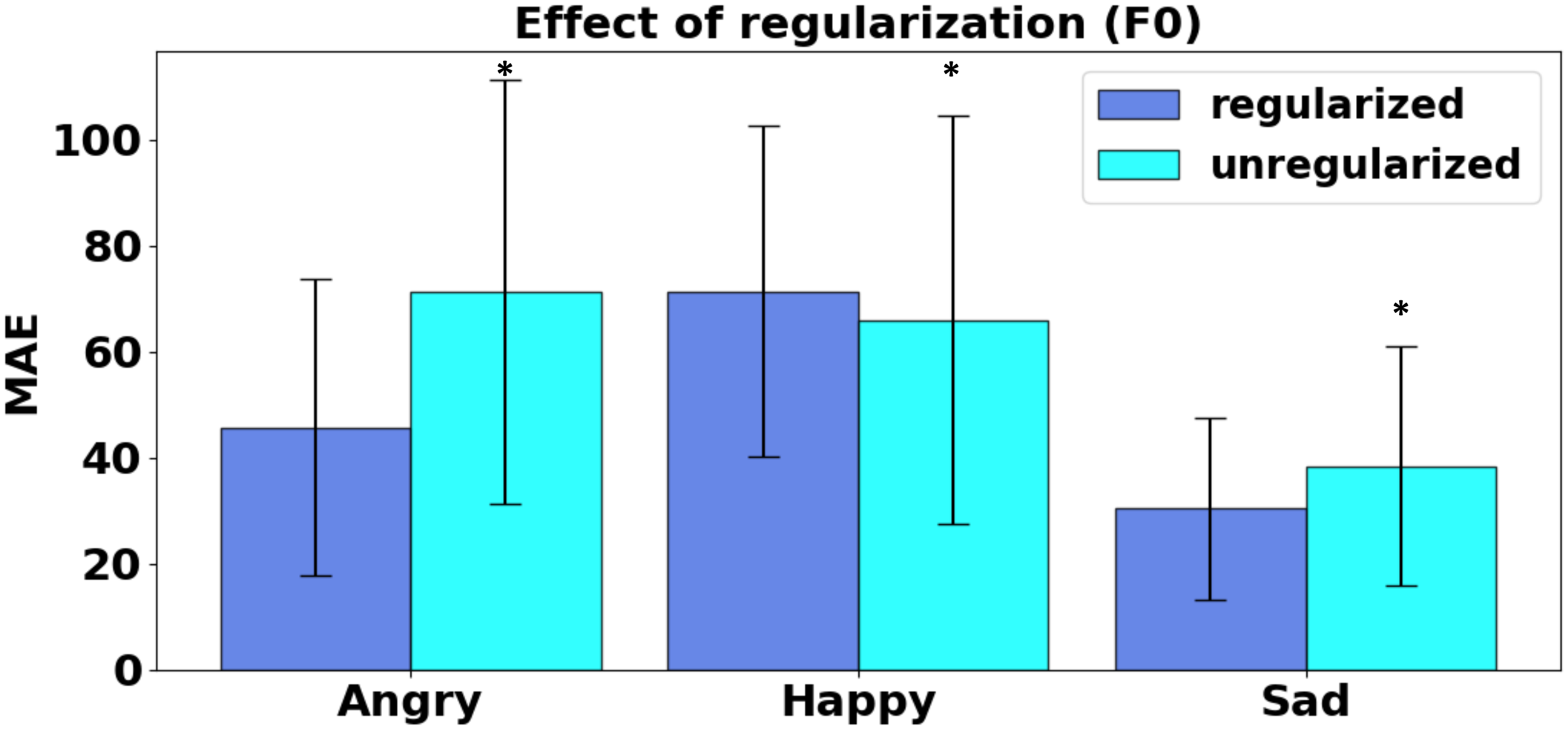}
  \caption{Effect of latent variable regularization on the prediction of fundamental frequency (F0) for each emotion pair. Marker $\**$ indicates $p<10^{-2}$ for paired t-test scores.}
  \label{fig:regularization_effect}
  \vspace{-1mm}
\end{figure}

\vspace{-1.5mm}
\subsection{Experimental Results}
\begin{figure}[!t]
  \centering
  \includegraphics[width=0.93\linewidth, height=6.3cm]{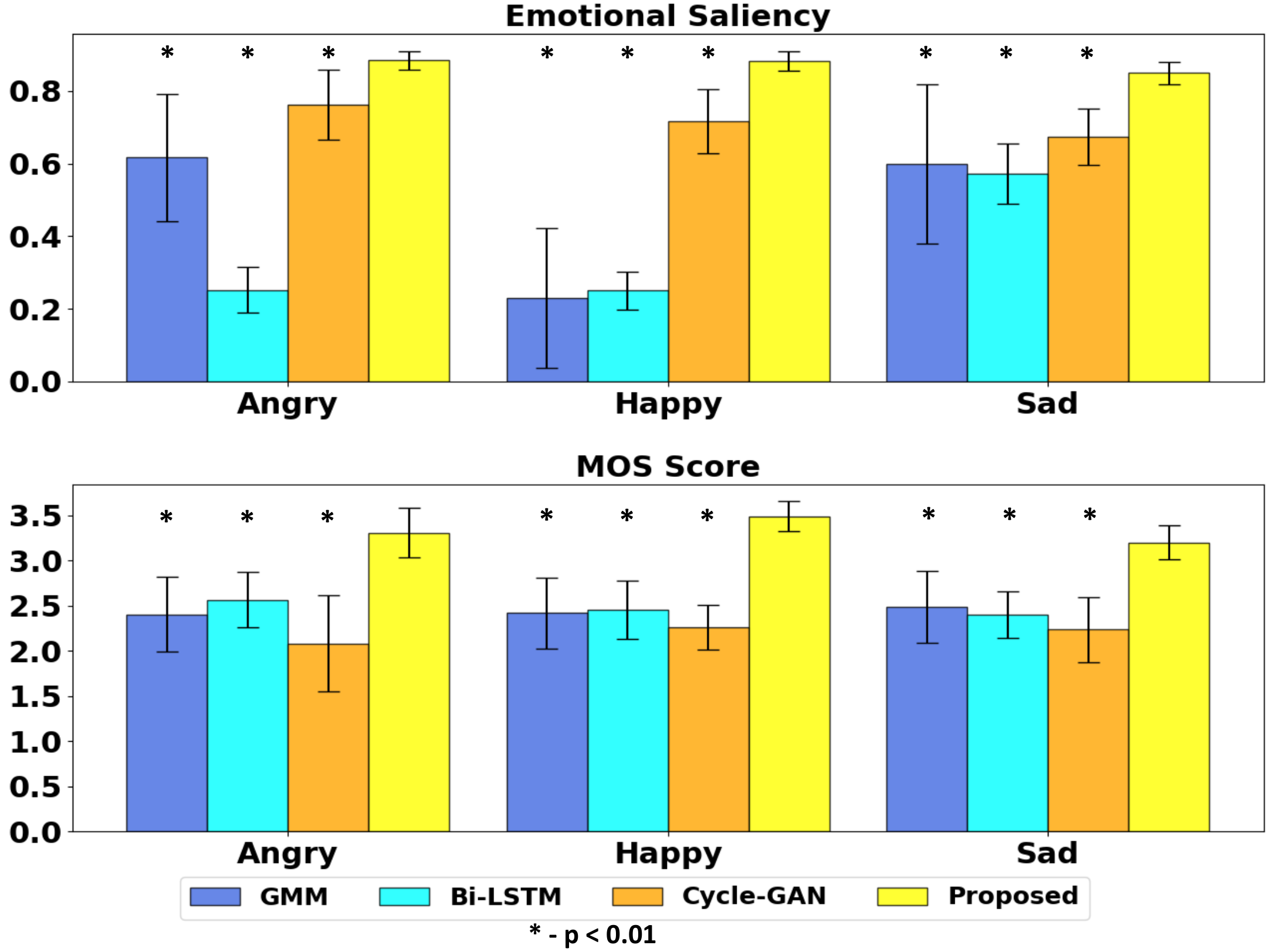}
  \caption{Confidence of emotion conversion (top) and the quality of reconstruction (bottom) for VESUS test samples.}
  \label{fig:vesus_results}
  \vspace{-6mm}
\end{figure}
As a sanity check, we carry out an ablation study to understand the effect of latent variable regularization via the LDDMM momenta. Fig.~\ref{fig:regularization_effect} shows the resulting mean absolute error in pitch prediction for each emotion pair. As seen, the F0 prediction is statistically significantly better in two emotional pairs. Neutral to happy conversion is an exception to this general trend, but we conjecture that this is due to the smaller training dataset ($\sim$800 samples compared to \textgreater 1400 for angry and sad). The error bars in all three emotion pairs are however, tighter than the un-regularized model, indicating that it is more robust.

\subsubsection{Mixed Speaker Evaluation}
Fig.~\ref{fig:vesus_results} illustrates crowd-sourcing results on the VESUS test dataset. Our proposed method has the highest emotional saliency rating in comparison to the baselines. The GMM did not produce intelligible speech when trained in a multi-speaker setting, as the F0 and spectral features do not exhibit distinct clusters when aggregated across speakers. Hence, the results in Fig.~\ref{fig:vesus_results} correspond to single-speaker training/testing. We note that our GMM evaluation is unfairly optimistic, and yet, the performance is worse than our method and the cycle-GAN. The Bi-LSTM model which simultaneously predicts the wavelet coefficients for F0 and energy, along with the spectrum has very poor conversion results for angry and happy. It is likely that the Bi-LSTM focuses on a subset of the features to minimize the overall loss. The cycle-GAN, on the other hand does produce reasonable results even though it is unsupervised. This is likely due to the implicit regularization produced by cyclic consistency and identity loss~\cite{cycle_gan}. Lastly, our proposed model has the best conversion score for all three emotion pairs and the tightest error bars in comparison to the baselines. Thus, our approach of combining the local and global task in a chained model works extremely well by allowing the individual pieces to train efficiently without losing oversight of the end goal. 
\par
The bottom plot in Figure~\ref{fig:vesus_results} shows the subjective quality of speech reconstruction after emotion conversion measured using mean opinion score (MOS). The chained neural network is uniformly better than the baseline algorithms on the VESUS dataset. It means that the proposed approach not only converts the emotion with a high degree of confidence but also manages to keep the quality of speech intact after conversion. 

\vspace{-1mm}
\subsubsection{Out-of-Sample Generalization}
\begin{figure}[!t]
  \centering
  \includegraphics[width=0.97\linewidth, height=6.3cm]{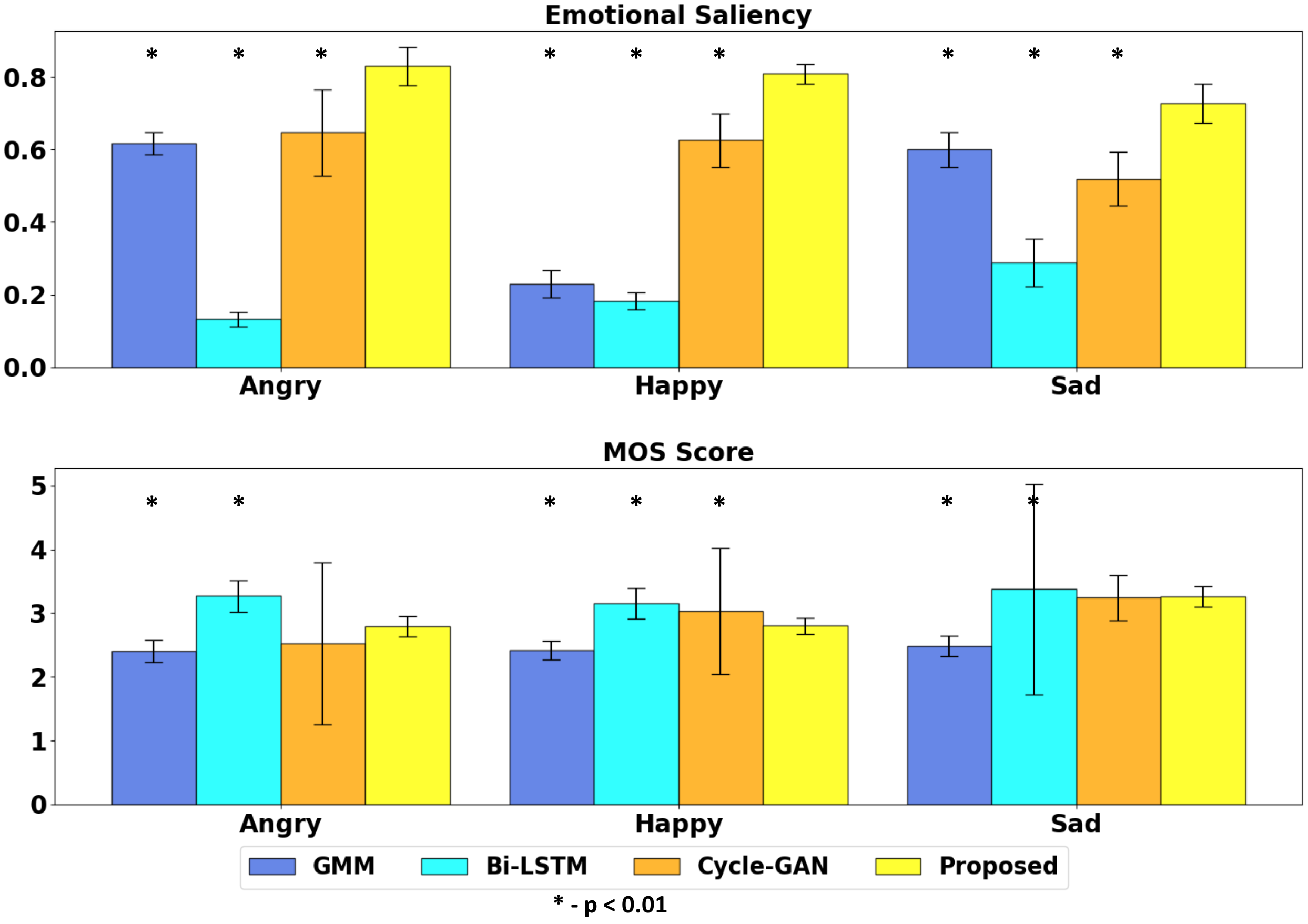}
  \caption{Confidence of emotion conversion (top) and the quality of reconstruction (bottom) on unseen samples.}
  \label{fig:oos_results}
  \vspace{-6.5mm}
\end{figure}
We further conduct an out-of-vocabulary emotion conversion experiment. Here, we set aside 7 randomly selected phrases per speaker from each emotion category. These phrases are not part of the training set to simulate unseen utterances during testing. Fig.~\ref{fig:oos_results} shows the results of this experiment. The GMM results are based on single-speaker evaluation. Once again, the proposed model has the best conversion performance with narrow error bounds. The Bi-LSTM does worse on unseen utterances demonstrating a lack of generalization capability. On the other hand, the cycle-GAN degrades a little but the saliency stays above 0.5 for all three emotion pairs. This is mainly due to the non-parallel nature of the Cycle-GAN model which makes no assumption on the speakers or the utterances. Our approach achieves this by not normalizing the input features using cohort statistics. Taken together, conditioning the spectrum estimation on the pitch can learn a complex relationship between the two which can be efficiently exploited as in our case. 
\par
The MOS in Fig.~\ref{fig:oos_results} show that Bi-LSTM has the best quality of reconstruction among the three. Empirically, it does not modify the speech at all, thereby, making it sound more natural by default. There is a tie for the second place between Cycle-GAN and the proposed model. Our proposed approach has much smaller error bars than Cycle-GAN due to training with un-normalized features and momenta regularization.

\vspace{-2mm}
\section{Conclusions}
We have proposed a novel method for emotion conversion that modifies pitch and spectrum using a chained neural network. Our proposed approach used a latent variable to regularize the F0 estimation, which in turn affects the spectrum prediction. We showed that using a diffeomorphic prior on the F0 contour and conditioning of spectrum on it leads to better generalization on unseen utterances. The experiments were carried out on the VESUS dataset and results on converted test samples were statistically significant. We finally conclude that our proposed algorithm did not degrade the quality of speech during conversion, thereby, exhibiting its all-round performance.

\bibliographystyle{IEEEtran}

\bibliography{mybib}

\end{document}